\documentclass[prl,showpacs,twocolumn]{revtex4}

\usepackage{graphicx}
\usepackage{amsmath}
\usepackage{slashed}
\usepackage{feynmp}

\begin{document}
\title{On the Origin of Lepton and Quark Masses}
\author{Ji\v r\'{\i} Ho\v sek}
\email{hosek@ujf.cas.cz}
\author{J. Adam, Jr.}
\affiliation{Department of Theoretical
Physics, Nuclear Physics Institute, Czech Academy of Sciences, 25068
\v Re\v z (Prague), Czech Republic}

\begin{abstract}
Gauging the flavor (family, generation, horizontal) index of the
chiral fermion fields of the Standard model, for anomaly freedom
extended by three sterile right-handed neutrino fields, results in
asymptotically free, {\it bona fide} nonconfining $SU(3)_f$ quantum
flavor dynamics. Approximate nonperturbative strong-coupling
solutions of the corresponding Schwinger-Dyson (SD) equation for
fermion self-energies give rise to the complete flavor symmetry
breaking by : (1) Three huge Majorana masses of sterile right-handed
neutrinos. (2) Three exponentially light Dirac masses common to all
fermion sorts in a family. Masses of charged leptons and quarks are
further distinguished from Dirac neutrino masses by the weak
hypercharge contributions to the universal $SU(3)_f$ kernel of the
SD equation, free of unknown parameters. The $SU(3)_f$ dynamics
itself thus gives the neutrino mass spectrum in the seesaw form.

\end{abstract}


\maketitle

\section{I. Introduction}

Understanding the wide and wild mass spectrum of quantum fields of
neutrinos, charged leptons and quarks is an alluring challenge of
theoretical elementary particle physics. The valuable achievement of
recent past is its phenomenological, theoretically unobjectionable
parametrization: The charged lepton and quark masses are described
in the Standard model (SM) by the essentially classical Higgs
mechanism \cite{higgs}, and the extremely light neutrino masses are
described in its minimal extension by the entirely classical seesaw
\cite{seesaw}. With fermion mixing the Lagrangian contains about two
dozens of theoretically arbitrary parameters which differ by at
least twelve orders of magnitude. Such a state of affairs is
unsatisfactory \cite{lee}. According to the standard understanding
of the energy spectra of the genuine quantum systems like
oscillators, hadrons, nuclei, atoms and molecules, the mass spectrum
of quantum fields of leptons and quarks should also be {\it
calculable}.

Calculable mass spectrum of fermion fields viewed as coupled quantum
oscillators is conceivable \cite{pagels} by replacing the Higgs
mechanism by the dynamical gauge symmetry breakdown pioneered by
Nambu \cite{njl}. Necessity of dealing with non-perturbative
techniques at strong coupling requires, however, approximations
which {\it a priori} are not under control \cite{pagels}. The
attempt formulated in \cite{hosek} and illustrated here, following
the suggestion of Yanagida \cite{yanagida}, is no exception. We
believe that the obtained results might justify the used
approximations {\it a posteriori}.

The paper is structured as follows. In Sect.~II we briefly summarize
the main properties of the model described in detail in
\cite{hosek}. The necessary new strong dynamics introduced there is
the gauge quantum flavor dynamics of three SM chiral fermion
families extended for anomaly freedom by three sterile right-handed
neutrinos. Section~III demonstrates the universal fermion flavor
mass splitting fundamentally different for Majorana and Dirac
masses. This analysis fixes the neutrino mass spectrum uniquely in
the seesaw form \cite{hosek1}. In Sect.~IV we describe how the weak
hypercharge contributions to the $SU(3)_f$ kernel of the SD equation
can provide large mass splitting observed within each generation
between the charged leptons and quarks with different electric
charges. In Sect.~V we summarize how the strong-coupling quantum
flavor dynamics efficiently replaces the weakly coupled Higgs sector
of the Standard model, and provide an illustrative fit of the
fermion mass spectrum. Sect.~VI contains our brief conclusions.

\section{II. Quantum flavor dynamics}

Gauging the flavor (horizontal, family, generation) symmetry of SM
is so natural that it could hardly be new
\cite{gaugedfamilysymmetry}. In the present form the model is
defined by gauging the flavor $SU(3)_f$ triplet index of three
chiral SM lepton ($l_{fL}, e_{fR}$) and quark ($q_{fL}, u_{fR},
d_{fR}$) families of the $SU(2)_L \times U(1)_Y$ gauge invariant SM.
This amounts to introduction of the octet of gauge flavor gluons
$C_a^{\mu}$, and for anomaly freedom to addition of one triplet of
sterile right-handed neutrino fields $\nu_{fR}$. In \cite{yanagida}
this $SU(3)_f \times SU(2)_L \times U(1)_Y$ gauge symmetry is
spontaneously broken down to $U(1)_{em}$ by an extended sector of
elementary scalar Higgs fields. In \cite{hosek} we argue that no
Higgs fields are necessary, i.e., {\it the strong flavor dynamics
itself self-consistently completely self-breaks}. We believe this is
a particular realization of the Nambu's idea \cite{njl} of dynamical
gauge symmetry breaking. The resulting anomaly free, asymptotically
free gauge $SU(3)_f$ quantum flavor dynamics is characterized by one
parameter. It is either the dimensionless gauge coupling constant
$h$ or, due to the dimensional transmutation, the theoretically
arbitrary scale $\Lambda$. Because the new dynamics results from
gauging another
fermion index we follow the habit and suggest, as others did already
before, to call it  the 'quantum flavor dynamics' (QFD).

Both in QCD and in QFD all the left- and the right-handed fermion
fields transform as triplets of the gauge group $SU(3)$, i.e., their
Lagrangians are formally identical. Consequently, in perturbation
theory, i.e., at short distances, these two theories must be
identical. In particular, both are asymptotically free. Despite this
we believe \cite{hosek} that at the strong coupling they are
entirely different. The QCD confines all its colored oscillator-like
excitations, whereas the QFD self-consistently generates the masses
to all its flavored ones.

The difference is in different electric charges of fermion fields:
In QCD, dealing with the electrically charged quark fields, only the
Dirac mass terms are possible. The Lagrangian (hard) mass terms are
allowed by the color $SU(3)$ symmetry (the product $\bar 3 \times 3
= 1 + 8$ does contain unity), and the strong low-momentum QCD
corrections to them are harmless for the color confinement (generate
the constituent quark masses). In contrast, the QFD deals also with
the electrically neutral neutrinos. The $SU(3)_f$ invariant hard
Dirac mass term common to all fermion sorts is obviously also
allowed, but there is a generically new possibility of the effective
Majorana mass of the neutrinos at large distances. For the
right-handed ones it has the form
\begin{equation*}
{\cal L}_{Majorana}= -\tfrac{1}{2}(\bar \nu_{R}M_{R}
(\nu_{R})^{{\cal C}}+ h.c. ) \ . \label{majorana}
\end{equation*}
It is, however, strictly prohibited  by the flavor $SU(3)$ symmetry:
The product $\bar 3 \times \bar 3 = 3 + \bar 6$ does not contain
unity. {\it As the Dirac mass term also the Majorana mass term
connects the right- and the left-handed fermion fields}: The
charge-conjugate $(\nu_{R})^{{\cal C}}$ is a {\it left-handed
field}. Unlike the Dirac mass term the left-handed charge-conjugate
right-handed neutrino field transforms as an {\it antitriplet} of
$SU(3)_f$. {\it Consequently, the QFD is not vector-like as QCD, but
at strong coupling it is effectively chiral}. The strong QFD quantum
corrections, if energetically favorable, generate the Majorana
masses dynamically, and this has the far reaching consequence: All
eight flavor gluons acquire masses by absorbing eight composite
'would-be' Nambu-Goldstone bosons as their longitudinal polarization
states \cite{jj}, and QFD gets completely self-broken.

Ultimately, also the Dirac masses of all fermions have to be
generated dynamically. The point is that {\it all hard QCD and QFD
Dirac mass terms are strictly prohibited by the chiral gauge
electroweak $SU(2)_L \times U(1)_Y$ interactions always present in
the game} at least as weak external perturbations.

Lack of systematic analytic strong coupling methods both in QCD and
QFD implies that for the description of majority of the
nonperturbative low-energy phenomena we are sentenced to using
models and rough approximations. In QCD the systematic analytic
method is the chiral perturbation theory \cite{heiri}, and the
systematic numerical method is the lattice \cite{latticeqcd}. In the
effectively chiral QFD we are not aware of any systematic analytic
method, and the lattice methods apparently do not apply
\cite{kaplan}.

\section{III. Neutrino seesaw mass matrix}

Following the suggestion of T. Yanagida we have demonstrated in
\cite{hosek} that in the anomaly free gauged three-flavor $SU(3)_f
\times SU(2)_L \times U(1)_Y$ model no Higgs fields are needed. {\it
Strong flavor gluon interactions themselves}, treated in a separable
approximation, clearly distinguish between the Majorana and the Dirac
masses. They result in the huge Majorana masses of sterile
neutrinos, and in naturally light hierarchically split Dirac masses
of the electroweakly interacting leptons and quarks.

This can be understood by looking at the Fig.1 of \cite{hosek1} as
follows:

In flavor space the Majorana mass term transforms in general as
\begin{equation}
\bar 3 \times \bar 3 = 3_a + \bar 6_s \ ,
\end{equation}
where the subscripts abbreviate the antisymmetric (a) and symmetric
(s) representations. Because the right-handed neutrino fields are
sterile, the Pauli principle uniquely selects the symmetric sextet.

In flavor space the Dirac mass term transforms differently:
\begin{equation}
\bar 3 \times 3 = 1 + 8 \ .
\end{equation}
The difference between $\bar 3 \times \bar 3$ and $\bar 3 \times 3$
translates into different combinations of the effective low-energy
parameters $g_{ab}$ which determine $M_R$ and $m_D$ in the
Schwinger-Dyson equation.

Three Majorana masses of the right-handed neutrinos come out huge,
of order $\Lambda$
\begin{equation}
M_{fM} \sim \Lambda \ . \label{MM}
\end{equation}
The neutrinos, charged leptons and quarks  of three generations
acquire the universal Dirac masses
\begin{equation}
m_{fD}=\Lambda \phantom{b} \rm exp \phantom{b} (-1/4\alpha_f) \ ,
\label{mD}
\end{equation}
where
\begin{eqnarray*}
\alpha_{1}&=&\tfrac{3}{64\pi^2}(g_{33}+\tfrac{2}{\sqrt
3}g_{38}+\tfrac{1}{3}g_{88}) \ , \\
\alpha_{2}&=&\tfrac{3}{64\pi^2}(g_{33}-\tfrac{2}{\sqrt
3}g_{38}+\tfrac{1}{3}g_{88}) \ , \\
\alpha_{3}&=&\tfrac{3}{64\pi^2}\tfrac{4}{3}g_{88} \ ,
\end{eqnarray*}
form the real diagonal matrix $\alpha$ in flavor space. It can be
expanded as
\begin{equation}
\alpha=\tfrac{3}{64\pi^2}[a 1 + b\lambda_3 + c\lambda_8]  \ ,
\label{alpha}
\end{equation}
where
\begin{eqnarray*}
a&=&\tfrac{2}{3}(g_{33}+g_{88})  \ , \\
b&=&\tfrac{2}{\sqrt 3}g_{38}  \ , \\
c&=&\tfrac{1}{\sqrt 3}(g_{33}-g_{88})  \ .
\end{eqnarray*}
Because for the neutrinos there are no other contributions to the
mass matrices $M_M$ a $m_D$, they combine into the famous seesaw $6
\times 6$ symmetric mass matrix \cite{seesaw}
\begin{equation}
\left(\begin{array}{ccc}0 & m_D\\ m_D^T & M_M\\ \end{array} \right) \ . \label{seesaw}
\end{equation}
After diagonalizing it describes three Majorana neutrinos with huge
masses $M_{\nu} \sim \Lambda$, and three extremely light Majorana
neutrinos with masses $m_{\nu} \sim m_D^2/M_M$.

The scale $\Lambda$ is theoretically arbitrary and must be fixed
once for ever from one appropriately chosen experimental datum. In
any case it is huge, because the flavor gauge symmetry, if it is
real, is badly broken and yet unobserved.

Natural possibility is to relate the new mass scale $\Lambda$ with
the nonzero neutrino masses. The trouble is that their values are
not known. From the experimentally available constraints the
preferred value is around $\Lambda \sim 10^{14}\, $GeV. In
\cite{hosek} we have argued that $\Lambda$ can be fixed also from
the invisibility of one particular composite pseudo NG boson, the
QCD axion, the existence of which the QFD also implies. The
resulting numerical value of $\Lambda$ is similar.

If the hypothesis of the complete dynamical self-breaking of $SU(3)$
is correct there should be no theoretically arbitrary parameters in
$M_R$ and $m_D$ except $\Lambda$. The effective low-energy constants
$g_{ab}$ should be calculable in terms of the pure numbers, like the
invariant group characteristics, e.g. the Clebsch-Gordan
coefficients or Casimir operators of various $SU(3)$
representations. {\it Ultimately, the neutrino mass spectrum is
completely fixed by the strong $SU(3)_f$ dynamics.}

\section{IV. Masses of charged leptons and quarks}

Masses of the charged leptons and quarks are not fully determined by
the strong flavor gluon exchanges in the SD equation connecting the
right-(R) and the left-(L) handed fermion fields.

Both for the charged leptons and for the quarks there are also the
Abelian electroweak gauge field $B^{\mu}$ exchanges which differ by
different numerical values of weak hypercharges (the non-Abelian
$SU(2)$ electroweak gauge fields $A^{\mu}_i$ interact merely with
the left-handed fermion fields and do not contribute to their Dirac
masses):
\begin{eqnarray*}
Y(l_L)&=&-1,\phantom{bbb}Y(e_R)=-2,\phantom{bbb}Y(\nu_R)=0 \ , \\
Y(q_L)&=&\tfrac{1}{3},\phantom{bbbb}Y(u_R)=\tfrac{4}{3},\phantom{bbbbb}Y(d_R)=-\tfrac{2}{3}
\ .
\end{eqnarray*}
\noindent (1) For charged leptons the $RL$ $B$ exchange in the SD
equation is proportional to ${\bar
{g'}}^2(q)\tfrac{1}{4}Y(e_R)Y(l_L)$.

\noindent (2) For quarks with the charge $Q=2/3$ the $RL$ $B$
exchange in the SD equation is proportional to ${\bar
{g'}}^2(q)\tfrac{1}{4}Y(u_R)Y(q_L)$.

\noindent (3) For quarks with the charge $Q=-1/3$ the $RL$ $B$
exchange in the SD equation is proportional to ${\bar
{g'}}^2(q)\tfrac{1}{4}Y(d_R)Y(q_L)$.

\noindent (4) For neutrinos the $RL$ $B$ exchange in the SD equation
is proportional to ${\bar {g'}}^2(q)\tfrac{1}{4}Y(\nu_R)Y(l_L)$=0.

The electroweak $B$ interactions themselves, being weak all the way
up to the Planck scale cannot generate the fermion self energies
$\Sigma$ dynamically. But they definitely contribute to the full
$LR$ kernel of the SD equation for $\Sigma$ as described above. We
don't know at present how to incorporate convincingly these
contributions into the separable Ansatz. We expect that in the
non-perturbative solution these contributions become amplified by
the nonanalytic dependence of the result upon the effective
couplings.

It is utmost important that these extra contributions are {\it fully
determined}: The sliding coupling constant $\bar {g'}(q)$ is known
as are the values of the fermion weak hypercharges.

Another important property of the $B$ {\it interaction} is that it
does not feel flavor: It is identical for fermions of a given sort
in all three families. Since, however, the QFD provides the
universal mass splitting of three flavors the sliding coupling $\bar
{g'}(q)$ has to be considered together with the three values of $Y_R
Y_L$ at three values of momenta.

It is then natural to introduce the nine real parameters $g^i_f$,
where $i=e$ abbreviates the charged leptons, and ($f=1, 2, 3$ or $e,
\mu, \tau$); $i=u$ abbreviates the quarks with the charge $Q=2/3$,
and ($f=1, 2, 3$ or $u, c, t$);  $i=d$ abbreviates the quarks with
the charge $Q=-1/3$, and ($f=1, 2, 3$ or $d, s, b$). We assume that
these parameters modify phenomenologically the universal mass
formulas (\ref{mD}) by the electroweak $B$ contributions as follows:
\begin{equation}
m_f(i)=\Lambda \phantom{b} \rm exp \phantom{b} (-1/4\alpha_f(i)) \ ,
\label{m(i)}
\end{equation}
where
\begin{eqnarray*}
\alpha_{1}(i)&=&\tfrac{3}{64\pi^2}\{[g_{33}+\tfrac{2}{\sqrt
3}g_{38}+\tfrac{1}{3}g_{88}]+g^i_1\} \ , \\
\alpha_{2}(i)&=&\tfrac{3}{64\pi^2}\{[g_{33}-\tfrac{2}{\sqrt
3}g_{38}+\tfrac{1}{3}g_{88}]+g^i_2\} \ , \\
\alpha_{3}(i)&=&\tfrac{3}{64\pi^2}\{\tfrac{4}{3}g_{88}+g^i_3\} \ .
\end{eqnarray*}
This simple parametrization serves merely as a primitive
illustration of the more ambitious picture formulated in the
Abstract.


\section{V. Numerical illustration}

\subsection{1. General considerations}

First we summarize how the QFD at strong coupling serves as a
microscopic dynamics underlying the weakly coupled Higgs sector of
the Standard model:

I. Primary is the spontaneous, genuinely quantal generation of
different huge Majorana masses $M_{fR}\sim \Lambda$ of three sterile
right-handed neutrinos, and of three different exponentially small
Dirac masses $m_{fD}= \Lambda \phantom{b}\rm exp (-1/4\alpha_f)$
common to all fermions $\nu_f, e_f, u_f, d_f$ in a family. These
fermion masses break spontaneously the $SU(3)_f \times SU(2)_L
\times U(1)_Y$ symmetry down to unbroken $U(1)_{em}$.

II. The underlying Goldstone theorem has two, valuable and firm,
gold and stone, consequences:

First, eight 'would-be' NG bosons composed predominantly of sterile
neutrinos (six of them have an admixture of the SM fermion
composites) give self-consistently rise to different huge calculable
masses of all flavor gluons. Here the self-consistence means that
for the formation of the longitudinal spin states of flavor gluons
their strong dynamics is crucial. The flavor $SU(3)_f$ gauge
symmetry gets dynamically badly completely self-broken, and
practically all its consequences are indirect.

Second, three multi-component 'would-be' NG bosons composed by the
strong QFD of all SM fermions give rise to the calculable masses of
the electroweak gauge bosons $W$ and $Z$. Here the electroweak gauge
interactions do not play any dynamical role. They are treated merely
as weak external perturbations.

III. Masses of the $W,Z$ bosons \cite{hosek} are given in terms of
the universal Dirac masses $m_{fD}$. This implies that the Weinberg
relation $m_W/m_Z=\rm cos \phantom{b} \theta_W$ is exact. Saturation
of sum rules for $m_W, m_Z$ by the mass of the heaviest (third)
family implies
\begin{equation*}
m_W^2=\tfrac{1}{4}g^2\tfrac{5}{4\pi} m_{3D}^2 \ ,
\end{equation*}
and enables to fix the mass $m_{3D}$ as
\begin{equation}
\boxed{m_{3D}\doteq 390\, {\rm GeV}\ . }
 \label{m3D}
\end{equation}
Obviously, this is the directly unobservable heaviest Dirac neutrino
mass entering the famous seesaw mass formula.

IV. Inclusion of quantum effects of the electroweak $B$
interactions, weakly coupled at $\Lambda$, results in mass splitting
of charged leptons and quarks within a given family. Study of the
influence of this weak coupling effect on the masses of the
intermediate bosons $W$ and $Z$ requires extra work. Our experience
with modeling this effect \cite{hosek2} suggests that it should be
small.

V. The model predicts six massive composite $0^{+}$ particles
\cite{hosek} as the partners completing the sets of the composite
'would-be' NG bosons into a representation of the corresponding
gauge group: (1) There should be one flavorless Higgs-like boson $h$
completing three multi-component electroweak 'would-be' NG bosons
composed of the electroweakly interacting fermions into a composite
(complex) $SU(2)$ doublet. (2) There should be two flavored,
flavor-conserving spinless bosons $h_3$ and $h_8$ completing six
flavored components of the 'would-be' NG bosons composed of the
electroweakly interacting fermions into a composite real flavor
octet. (3) There should be three superheavy flavored spinless bosons
$\chi_i$ completing eight components of flavored 'would-be' NG
bosons composed of sterile right-handed neutrinos into the composite
complex flavor sextet $(2 \times 6 = 3 + 8 + 1)$ (one pseudo-NG
boson remains in the physical spectrum as one of three axions).

It is instructive to compare the steps above with the corresponding
steps in the weakly coupled Higgs sector of the Standard model:

ad I. Primary is to arrange the classical Higgs-field potential into
the form which allows for the spontaneous breakdown of the gauge
$SU(2)_L \times U(1)_Y$ symmetry down to $U(1)_{em}$.

ad II. The underlying Goldstone theorem has one valuable
consequence: Three elementary 'would-be' NG bosons, pre-prepared in
the complex Higgs field doublet become at the tree level the
longitudinal spin states of $W$ and $Z$ bosons. Their masses are
proportional to the Higgs-field condensate.

ad III. Because of the symmetry of the Higgs field kinetic term the
Weinberg relation is fulfilled.

ad IV. Inclusion of fermions, i.e. adding their invariant Yukawa
couplings with the Higgs field is an independent and fortunate step.
These couplings generate at tree level for free the theoretically
arbitrary fermion masses. Their quantum effects on the robust
tree-level generation of $m_W, m_Z$ are essentially negligible.

ad V. There is one elementary massive Higgs boson as a remnant of
the elementary complex $SU(2)$ doublet.

\subsection{2. Fitting the fermion mass spectrum}

1. We start by {\it assuming} that the heaviest Dirac neutrino mass
is $m_{3D}=390\, $GeV, fixed from the sum rule for
$W,Z$ masses. For definiteness we set $\Lambda = 10^{14}\, $ GeV.
From (\ref{mD}) we easily compute $g_{88}=1.502790\ .$

2. With the known $g_{88}$ and with the experimental values of
$m_{\tau}, m_t, m_b$ (TABLE I)  we fix $g_3^e =-0.341190$,
$g_3^u =-0.060098$, $g_3^d =-0.295027$.

3. From the experimental values of fermion masses of two lighter
families (TABLE I) we fix the right-hand sides of the equations
\begin{eqnarray*}
g_{33}+\tfrac{2}{\sqrt 3}g_{38}+g_1^e=a_1 \ , \\
g_{33}+\tfrac{2}{\sqrt3}g_{38}+g_1^u=b_1 \ , \\
g_{33}+\tfrac{2}{\sqrt 3}g_{38}+g_1^d=c_1 \ , \\
g_{33}-\tfrac{2}{\sqrt 3}g_{38}+g_2^e=a_2 \ , \\
g_{33}-\tfrac{2}{\sqrt3}g_{38}+g_2^u=b_2 \ , \\
g_{33}-\tfrac{2}{\sqrt 3}g_{38}+g_2^d=c_2 \ .
\end{eqnarray*}
These are the six linear inhomogeneous equations for eight unknown
parameters $ g_f^e, g_f^u, g_f^d, g_{33}, g_{38}$, $f=1,2$.

\begin{table}[t]
\caption{\label{tab1} Fermion masses from experiment \cite{PDG}.}
\begin{ruledtabular}
\begin{tabular}{ccc}
$m_3(e)\equiv m_\tau$  & $m_2(e)\equiv m_\mu$   &  $m_1(e)\equiv m_e$   \\
\hline
    $1.777\, $GeV        & $105.7\, $MeV         &  $0.511\, $MeV  \\ \hline \hline
$m_3(u)\equiv m_t$       & $m_2(u)\equiv m_c$    &  $m_1(u)\equiv m_u$   \\
\hline
    $173.1\, $GeV        & $1.28\, $GeV          &  $2.2\, $MeV         \\ \hline \hline
$m_3(d)\equiv m_b$       & $m_2(d)\equiv m_s$    &  $m_1(d)\equiv m_d$  \\
\hline
    $4.18\, $GeV         & $96.0\, $MeV          &  $4.6\, $MeV        \\
\end{tabular}
\end{ruledtabular}
\end{table}

4. Consequently, there is a two-parameter freedom in fixing $g_{33}$
and $g_{38}$, constrained by the assumption from the point 1:
$m_{iD}<m_{3D}$, $i=1,2$. Clearly, would we know $g_{33}$ and
$g_{38}$ from the neutrino mass spectrum given by the seesaw mass
formula, the system for the unknown $g_f^e, g_f^u, g_f^d$ would be
uniquely fixed.

5. For an illustration we fix the remaining universal parameters of
QFD as $g_{33}=1.27262$ and $g_{38}=-0.0594915$, corresponding to
$m_{1D}=3.9\, $GeV, $m_{2D}=39\, $GeV and $m_{3D}=390\, $GeV. The
resulting values of the weak hypercharge contributions to the
charged lepton and quark masses are collected in TABLE II. It is
gratifying that they all are essentially of the same order of
magnitude.

\begin{table}[b]
\caption{\label{tab1} The parameters of the theory describing
the fermion spectra.}
\begin{ruledtabular}
\begin{tabular}{ccc}
     $g_{33}$     & $g_{38}$     &  $g_{88}$    \\ \hline
      1.27262    & -0.0594915    &  1.502790  \\ \hline \hline
     $g_1^e$     & $g_2^e$       &  $g_3^e$    \\ \hline
      -0.382808  & -0.315776     &  -0.341190  \\ \hline \hline
     $g_1^u$     & $g_2^u$       &  $g_3^u$    \\ \hline
    -0.332490    & -0.196766     &  -0.060098  \\ \hline \hline
     $g_1^d$     & $g_2^d$       &  $g_3^d$    \\ \hline
    -0.305581    & -0.320025     &  -0.295027   \\
\end{tabular}
\end{ruledtabular}
\end{table}

With these illustrative numbers the masses of three active Majorana
neutrinos can be roughly estimated, ignoring the matrix structure of
seesaw, as $m_{\nu_{\tau}}\sim m_{3D}^2/\Lambda\doteq
1.521\phantom{b}\rm eV$, $m_{\nu_{\mu}}\sim m_{2D}^2/\Lambda\doteq
1.521\times 10^{-2}\phantom{b} \rm eV$, $m_{\nu_e}\sim
m_{1D}^2/\Lambda\doteq 1.521\times 10^{-4}\phantom{b} \rm eV$.

6. It follows from the explicit illustration presented above that
the fermion masses are related with each other in a rather
sophisticated way. First, six neutrino masses come from two QFD
sources: $M_{fR}$ and $m_{fD}$. Second, masses of the charged
leptons and quarks are described in terms of the QFD parameters of
$m_{fD}$, and in terms of the parameters associated with the weak
hypercharge.

{\it Ultimately}, however, the $SU(3)_f \times SU(2)_L \times
U(1)_Y$ gauge dynamics of the system of the chiral fermion fields of
the Standard model extended by three sterile right-handed neutrino
fields gives rise both to the gauge boson and the fermion masses
calculable solely in terms of $\Lambda$ and the electroweak
couplings.

\section{IV. Conclusion}

The picture painted in this paper is based on a very strong
assumption: The $SU(3)_f$ dynamics apparently identical with the QCD
dynamics does not confine  at large distances its elementary
constituents, but rather self-consistently generates the calculable
masses to all of them. Crucial is the possibility of generating the
Majorana neutrino masses. The separable Ansatz used for illustrating
this, though physically well motivated by the BCS \cite{njl,
fetter}, is not under theoretical control. We tend to defend
ourselves by F. Wilczek's \cite{wilczek} Jesuit credo: "It is more
blessed to ask forgiveness than permission." We justify our
perseverance by a number of desirable physical phenomena which the
present rigid model describes and correlates \cite{hosek}. Above
all, to convert the innocent fermion flavor index into the source of
a new force resulting in the universal hierarchical flavor splitting
is irresistibly suggestive. Suggestive is also to associate the mass
splitting within one family with the known electroweak force
\cite{huang}.

Gauging the family (flavor, horizontal, generation) index promises a
hint to a solid theoretical answer to the famous question 'why three
families' \cite{yanagida2}. The LEP experimental proof of the
existence of three light neutrinos \cite{lep} makes, of course, the
question rather academic. Observation of the Higgs-like scalars
$h_3$ and $h_8$ with calculable properties \cite{benes-hosek}, which
are the clear signature of the dynamical $SU(3)_f$ family picture
\cite{hosek} would be quite intriguing.
\\
\begin{acknowledgements}
The work on this project has been supported by the grant
LG 15052 of the Ministry of Education of the Czech Republic.
\end{acknowledgements}

\end{document}